# Organic-Inorganic Hybrid $CH_3NH_3PbI_3$ Perovskite Solar Cell Nanoclusters: Revealing Ultra-Strong Hydrogen Bonding and Mulliken Inner Complexes and Their Implication in Materials Design

Arpita Varadwaj,[*,a,b] Pradeep R. Varadwaj,[a,b] Koichi Yamashita[a,b]

[a]Department of Chemical System Engineering, School of Engineering, The University of Tokyo 7-3-1, Hongo, Bunkyo-ku, Japan 113-8656
[b]CREST-JST, 7 Gobancho, Chiyoda-ku, Tokyo, Japan 102-0076

Abstract

Methylammonium lead iodide ($CH_3NH_3PbI_3$) perovskite solar cell has produced a remarkable breakthrough in the photovoltaic history of solar cell technology because of its outstanding device based performance as a light-harvesting semiconductor. Whereas the experimental and theoretical studies of this system in the solid state have been numerously reported in the last 4 years, its fundamental cluster physics is yet to be exploited. To this end, this study has performed theoretical investigations using DFT-M06-2X/ADZP to examine the principal geometrical, electronic, topological, and orbital properties of the $CH_3NH_3PbI_3$ nanocluster blocks. These clusters are found to be unusually strongly bound, with binding energies lying between –93.53 and –125.11 kcal mol$^{-1}$ (beyond the covalent limit, –40 kcal mol$^{-1}$), enabling us to characterize the underlying interactions as *ultra-strong* type. Based on this, together with the unusually high charge transfers, strong hyperconjugative interactions, sophisticated topologies of the charge density, and short intermolecular distances uncovered, we have characterized the $CH_3NH_3PbI_3$ as Mulliken inner complexes. Additionally, the consequences of these, as well as of the *ultra-strong* interactions, in designing novel functional nanomaterials are briefly discussed. The various new results obtained in this study are not in perfect agreement with those already reported experimentally (Nat. Commun. 2015, 6, 7124), and computationally (Chem. Commun., 2015, 51, 6434; Sci. Rep. 2016, doi:10.1038/srep21687; Chem. Mater. 2016, 28, 4259; J. Mat. Chem A 2016, 4, 4728; J. Phys. Chem. Lett. 2016, 7, 1596).

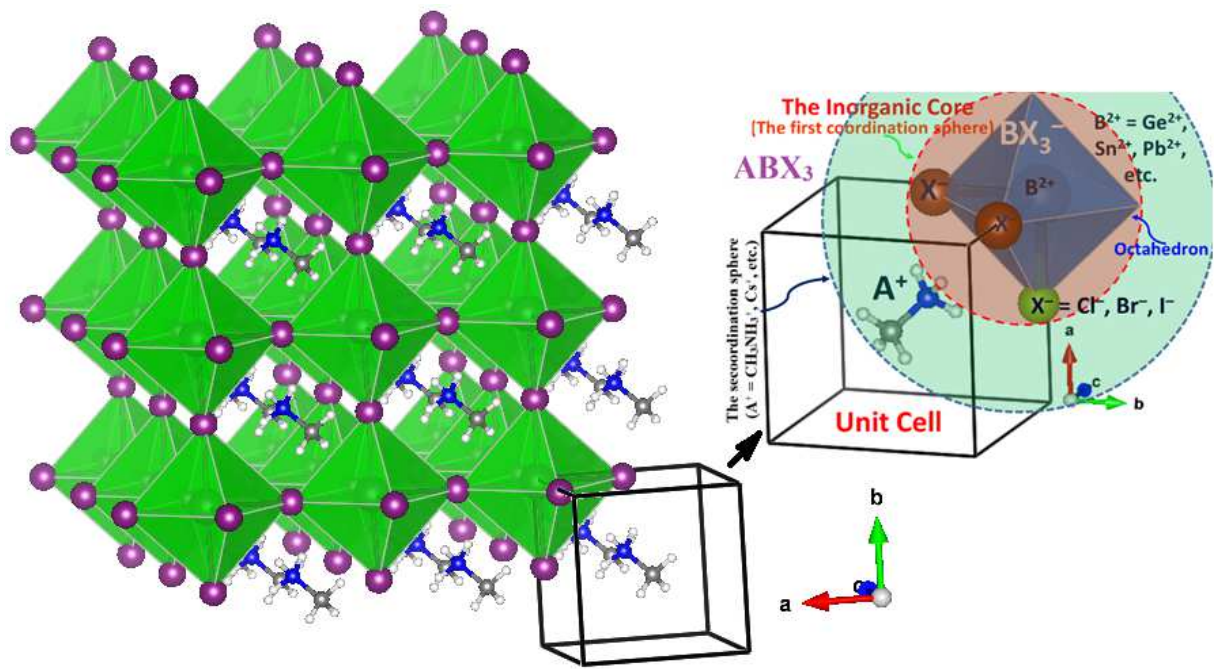

[*]Corresponding Author's E-mail Addresses: pradeep@t.okayama-u.ac.jp (PRV); varadwaj.arpita@gmail.com (AV); yamashita@chemsys.t.u-tokyo.ac.jp (KY)

## 1. Introduction

Designing high-performance photoresponsive visible-light-driven materials for solar cell-based future technologies is cutting-edge research.[1] Among others,[2] organic–inorganic lead/tin/germanium trihalide-based hybrid perovskites are a special class of novel materials that have created historical breakthroughs in photovoltaic technology.[3] Even though perovskite materials have been known to the scientific community for over a century,[4] they have only recently evoked renewed interest.[1] This was made possible by Kojima et al.,[5] who reported the photon-to-electricity power conversion efficiency (PCE) of the methylammonium lead(II) tribromide ($CH_3NH_3PbBr_3$) perovskite as 2.2% in 2006,[5a] and of methylammonium lead(II) triiodide ($CH_3NH_3PbI_3$) perovskite as 3.8% in 2009.[5b] These two fundamental studies were sufficient to inspire the solar energy community to carry out further research on these,[1,6] as well as on analogous systems,[7] for further development of the PCE value of these materials with variable experimental settings.[6a),b)] Over 2000 articles have already been published addressing various aspects of the trihalide-based perovskite solar cell materials. Interestingly, this was achieved within the very short period between 2012 and 2017,[8] and there are preparations of many of them that are currently on-going by diverse research groups. Although diverse halide-based perovskite solar cell systems have been experimentally synthesized, many of them are reported to be in two and three dimensions.[9] However, the $CH_3NH_3PbI_3$ perovskite solar cell is the only system that has shown very high semiconducting performance due to a number of its recommended characteristics.[10a] Some of them, for example, include very high light absorbing potential, giant photoinduced dielectric constant,[10b] very small excitation binding energy of only a few millielectronvolts (<<16 meV) at room temperature,[10c] very small excitonic reduced effective mass, ≈ $0.104m_e$ (where $m_e$ is the electron mass),[10c] and large effective diffusion lengths, about 100 nm for both electrons and holes (meaning large diffusion lengths for excitons as well). [1b),10e)–f)] Because of these specific features, the $CH_3NH_3PbI_3$ perovskite solar cell has been seriously focused on by a wide range of scientists,[1–10,11] leading to a National Renewable Energy Laboratory (NREL) certified PCE value of 22.1% reported early last year.[3,12]

Discovery of air- and moisture-stable perovskite solar cell materials with long-term device stability is one of the main concerns of the current solar cell research.[13] Therefore, serious attention is given to developing low-bandgap trihalide-based photovoltaic perovskite materials, which are of different varieties.[14] This is unsurprising because materials of these kinds have proven beneficial for the development of powerful ambipolar carrier transport devices,[10d] e.g., transistors and light-emitting diodes. These devices work both in accumulation (p-type) and inversion (n-type) modes.[15a] Another distinct advantage of these kinds of single-crystalline trihalide perovskite materials is that the maximum efficiency of photon conversion into electrical energy can boost photovoltaic device PCE up to 25%, or even up to 35%.[15b] Quan et al. have claimed that they were the first to report certified hysteresis-free solar power conversion in a planar perovskite solar cell,

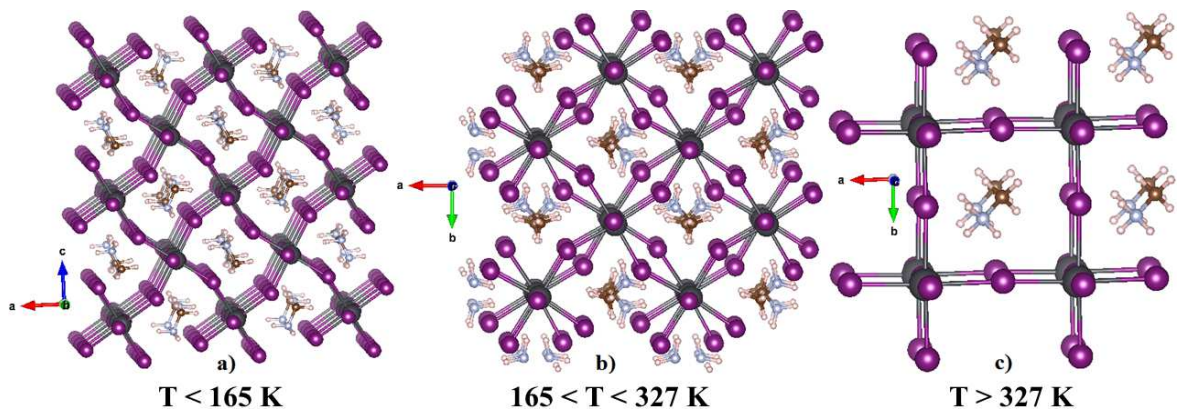

Scheme 1: The 2×2×2 models representing the a) orthorhombic, b) tetragonal, and c) pseudocubic geometries of the organic-inorganic $CH_3NH_3PbI_3$ perovskite solar cell. The values of temperature (T) are shown to signify stability (light-pink). Shown in the the ball and stick models are the iodine (purple), carbon (brown), nitrogen (light-blue), and hydrogen atoms. The views are presented along ac and ab planes for a) and b)–c), respectively.

in which the certified PCE was about 15.3%.[15c]

TriHalide perovskites have the general chemical formula $ABX_3$, where A is an inorganic/organic cation with +1/+2 oxidation state (e.g., $Cs^+$/$CH_3NH_3^+$/$NH_3NH_3^{2+}$), B is a metal ion with +2/+3/+4 oxidation state (e.g., $Pb^{2+}$/$Bi^{3+}$/$Sn^{4+}$), and X is a halogen derivative with an oxidation state of –1 (namely $Cl^-$/$Br^-$/$I^-$). The formula above is derived from $CaTiO_3$, a mineral that was discovered by Gustov Rose in 1839, and honors the Russian mineralogist Lev Aleksevich von Perovski.[16a] This is the reason why compounds such as $SrTiO_3$, $BiFeO_3$, and $Pb[Zr_xTi_{1-x}]O_3$ ($0 \leq x \leq 1$), etc., having geometrical topologies identical to $CaTiO_3$ are called perovskites.[16b] When the halogen derivative X in the formula $ABX_3$ refers to the monovalent iodide anion $I^-$, the symbol B to the $Pb^{2+}$ divalent cation, and A to the methylammonium $CH_3NH_3^+$ monovalent organic cation, the resulting species is methylammonium lead(II) triiodide ($CH_3NH_3PbI_3$), an organic–inorganic hybrid perovskite.[16c),d] It is experimentally known to be structurally viable in different temperature phases, such as orthorhombic, tetragonal, pseudocubic, and monoclinic (see Scheme 1 for the geometries of the former three).[16e),f)]

It is recently demonstrated that the $CH_3NH_3PbI_3$ perovskite solar cells with appropriate tandem technology are efficient for photocatalytic water splitting.[17] In addition, the bandgap of this kind of perovskite materials is determined by the supramolecular dimension. For example, the dimensional reduction of the perovskite material passing from 3D to 2D causes the bandgaps of the materials to increase. Chang et al. showed that the 2D quantum well perovskite structure has a bandgap of 1.86 eV, which is larger than that of its corresponding 3D analogue (1.17 eV).[18] Similarly, increasing the number of the lead-halide layers in the 2D quantum well structure decreases the bandgaps from 1.86 eV (one layer) through 1.59 eV (two layers) to 1.44 eV (three layers). This feature was suggested to be due to the carrier confinement in the lead-based halide perovskites unperturbed by the organic cation.

The present study is mainly focussed on the $CH_3NH_3PbI_3$ perovskite bulk when it is visualized as a supermolecule. It is doable only when the bulk is seen as a binary complex/cluster (in zero-dimension), which is possible when no periodic boundary condition is employed to it. The characteristic features of the cluster may in some way be similar to what might be expected for other binary clusters.[19] The main difference between these latter and the $CH_3NH_3PbI_3$ perovskite cluster lies in the fact that the latter is electrically neutral analogously as $Na^+Cl^-$ (wherein charge-charge type Coulombic interaction plays a major role), while the stability of the former ones may or may not have the same electrostatic origin.[19e)]

The $CH_3NH_3PbI_3$ supermolecular nanocluster may be assumed to be a by-product of two subunits, the $PbI_3^-$ inorganic anion (i.e., the triiodoplumbate(II) ion) and the $CH_3NH_3^+$ organic cation, driven by electrostatics. The attractive intermolecular interaction between the monomers forming this cluster is not unusual because the triangular face formed by the three $I^-$ ions in $PbI_3^-$ is electron rich, meaning it is highly nucleophilic, which is hence an ideal platform to accommodate the electron-deficient H atoms of the $-NH_3/-CH_3$ group in $CH_3NH_3^+$. This is perhaps the general understanding of many scientists, which has also appeared in many research articles.[20] While many researchers question whether the $CH_3NH_3^+$ cation plays any role,[20a)–d)] others add that its role is to make the perovskite crystal lattice electrically neutral,[20] as well as to provide the $CH_3NH_3PbI_3$ perovskite solar cell a shape after causing some structural modifications to the $PbI_3^-$ lattice.[20e)] Considering these remarkable observations as background, an important concern of this study is to investigate whether it is electrostatics alone, or whether there are other attractive factors (e.g., charge transfer) that must be considered to explain the stability of the $CH_3NH_3PbI_3$ nanocluster.

Fang and Jena recently considered the $CH_3NH_3PbX_3$ binary clusters, and some of the substituted derivatives that replace X by the superhalogens $BH_4^-$ or $HCOO^-$, $Pb^{2+}$ by $Sn^{2+}$ or $Ge^{2+}$, and $CH_3NH_3^+$ by $HC-(NH_2)_2^+$, as fundamental building blocks.[21] They concluded that the properties of the organic–inorganic hybrid perovskites such as the fundamental bandgap, the gap deformation potential, the exciton binding energy, and the hygroscopicity, all originate from their corresponding molecular moieties. In addition, they recommended adopting the simple molecular models that they employed, together with Goldschmidt's tolerance factor as an important combination, to make quick estimations for the stability and functional properties of new hybrid perovskite materials, which can serve as a preliminary screening process before accurate bulk calculations.

This study is fully focused for the first time to investigate theoretically the characteristic physical electronic properties of the parent $CH_3NH_3PbI_3$ nanocluster, and its possible conformers using density functional theory (DFT).[22] Specifically, the primary interest is to understand if the $CH_3NH_3PbI_3$ perovskite nanoclusters are stable in the gas phase (which may somehow be the vapor phase equivalent), and whether there is any connection of their gas phase properties with those of their solid-state properties, the latter evaluated using periodic DFT calculations. The secondary interest is to uncover the binding strengths of the intermolecular interactions formed between the $PbI_3^-$ and $CH_3NH_3^+$ subunits and to examine whether these strengths can be comparable with intermolecular interactions of any kind viable in the wide literature as strong and medium-to-weak strength hydrogen bonding interactions and van der Waals.[23] One of the most important concerns toward this end is to investigate whether intermolecular interactions in the $CH_3NH_3PbI_3$ nanoclusters are involved in any kind of proton-transfer features? Can such interactions be regarded as purely electrostatic, or can one recognize them as largely covalent, or a mixture of the two? A tertiary interest is to employ various state-of-the-art modeling tools (e.g., Natural Bond Orbital (NBO)'s second-order hyperconjugative analysis,[24] the quantum theory of atoms in molecules (QTAIM) topological analysis,[25] and the reduced density gradient noncovalent interaction (RDG-NCI) analysis[26]) to examine and to unravel the nature of the various chemical bonding interactions unitizing the monomeric fragments $PbI_3^-$ and $CH_3NH_3^+$ into clustered configurations, for which, they are becoming functional in the solid state in their bulk forms.

## 2. Computational details

The Gaussian 09 code[27a)] was utilized for the calculations of the geometries and electronic properties of the most stable $CH_3NH_3PbI_3$ cluster, and its possible conformations. The DFT-M06-2X[28] functional, in conjunction with the double-ζ DZP basis set,[29] was employed to calculate the first derivative of the energy with respect to the atom fixed nuclear coordinates, and one-electron properties of all the clusters examined. The choice of the DFT functional is obviously due to the fact that it is recommended as a global hybrid, and that it is suitable for applications involving main-group thermochemistry, kinetics, noncovalent interactions, and electronic excitation energies to valence and Rydberg states.[28]

The Hessian second derivative of the electronic energy with respect to the atom fixed nuclear coordinates were performed on each of the optimized geometries of the $CH_3NH_3PbI_3$ to insight into the nature of the geometries of the clusters identified. In all cases, the eigenvalues of the Hessian second derivative matrix were positive, meaning they are all minimum energy geometries with respect to the current level of theory.

It is worth mentioning that in most of the previous studies[30] the identification of the intermolecular noncovalent interactions in the various geometries of the lead triiodide perovskite solar cell in the three important phases of temperature was accomplished using intermolecular distance as a metric. However, as demonstrated previously on several occasions, characterizing this attribute only through the distance criterion is not standalone, as is not suitable alone for identifying interactions that are weak.[31,36] To this end, we have applied for the first time the QTAIM approach[25a)] implemented in AIMAll[25b)] to the $CH_3NH_3PbI_3$ clusters for insight into the various chemical bonding interaction topologies involved to unite the monomeric fragments $PbI_3^-$ and $CH_3NH_3^+$ in clustered configurations. This approach also allows us to quantify the interaction strengths in terms of charge density and other topological descriptors, including for example, the Laplacian of the charge density, the kinetic ($G_b$), potential ($V_b$), and total energy ($H_b = G_b + V_b$) densities. The delocalization index δ between the various atomic basin pairs are also evaluated with the framework of this theory. The RDG-NCI approach[26] proposed recently is employed to examine the extent to which it recovers the topological bonding interactions identified and characterized with QTAIM,[25a] which is indeed essential to establish further the efficacy of the latter method in unraveling the genuineness of the physical chemistry it presents to the scientific community.

Geometry relaxation and bandgap calculations using periodic DFT were performed using VASP.[27b]–[e] The Perdew–Burke–Ernzerhof (PBE) exchange-correlation functional, together with the projector augmented wave (PAW) potentials for all atoms, the cut-off energy for the plane wave basis set of 520 eV, and a 6 × 6 × 6 Γ-centered $k$-point sampling of the Brillouin zone were set for the periodic relaxation of the geometry of the $CH_3NH_3PbI_3$ in the high temperature pseudocubic phase, as well as for bandgap calculations. The tetrahedron method with Blöchl corrections for the Brillouin-zone integrations was utilized for this purpose.[27f]

In addition, we employed the NBO approach[24] to examine whether the physical chemistry obtained with the aforesaid charge density approaches can be compared with those gleaned from the natural orbital method. Specifically, our interest was to examine if the reliabilities of the bond paths developed between the various atomic basins constituting the triiodide perovskite clusters have any consistency with the second-order perturbative estimates of donor–acceptor (bond–antibond) interactions in the NBO basis, which are consequences of charge-transfer delocalization (hyperconjugation).[24a] According to this method, for a given set of donor NBO ($i$) and acceptor NBO ($j$), the stabilization energy $E^{(2)}$ for a given interaction type associated with delocalization $i \rightarrow j$ can be estimated with Eqn 1, where $q_i$ is the donor orbital occupancy, $\varepsilon_i$ and $\varepsilon_j$ are orbital energies (diagonal elements), and $F(i, j)$ is the off-diagonal NBO Fock matrix element.[24a]

$$E^{(2)} = \Delta E_{ij} = q_i \frac{|F(i,j)|^2}{\varepsilon_j - \varepsilon_i} \quad \ldots\ldots\ldots\ldots\ldots\ldots\ldots\ldots\ldots\ldots\ldots\ldots\ldots\ldots\ldots\ldots\ldots\ldots..1)$$

For the calculation of $\Delta E$, the sum of the total electronic energies of the two monomeric fragments $CH_3NH_3^+$ and $PbI_3^-$ ( $\sum_{Monomers} E_T(CH_3NH_3^+, PbI_3^-)$ ) was substracted from the total electronic energy $E_T(CH_3NH_3^+ \bullet\bullet\bullet PbI_3^-)$ of the given $CH_3NH_3PbI_3$ conformer (Eqn. 2)). The same approach was endorsed for calculating the change in the Gibb's free energy $\Delta G^0$, but in this case the sum of the total Gibb's free energies of the two monomers ( $\sum_{Monomers} G(CH_3NH_3^+, PbI_3^-)$ ) was substracted from the total Gibb's free energy $G_T(CH_3NH_3^+ \bullet\bullet\bullet PbI_3^-)$ ) of the given

CH$_3$NH$_3$PbI$_3$ conformer (Eqn. 3)). No basis set superposition error is considered since the basis sets used to study these systems undertaken with a basis set of double-ζ quality.

$$\Delta E = E_T(CH_3NH_3^+ \bullet\bullet\bullet PbI_3^-) - \sum_{Monomers} E_T(CH_3NH_3^+, PbI_3^-) \quad \text{............................2)}$$

$$\Delta G^0 = G_T(CH_3NH_3^+ \bullet\bullet\bullet PbI_3^-) - \sum_{Monomers} G(CH_3NH_3^+, PbI_3^-) \quad \text{............................3)}$$

Unless otherwise mentioned, and from now on, the methylammonium cation will be called as organic cation, molecular cation, CH$_3$NH$_3^+$, or MA$^+$, while the methylammonium lead trihalide perovskite will be referred as lead triiodide perovskite, CH$_3$NH$_3$PbI$_3$, CH$_3$NH$_3^+\bullet\bullet\bullet^-$PbI$_3$, or CH$_3$NH$_3^+\bullet\bullet\bullet^-$I$_3$Pb. The binding and delocalization energies will be discussed in units of kcal mol$^{-1}$, while bandgap in eV.

### 3. Results and Discussion
#### 3.1 Structural and energetic stabilities, and potential energy surface

As already indicated in the Introduction, one of the most important outlooks of significant general interest is whether the CH$_3$NH$_3$PbI$_3$ supramolecular nanocluster is stable in the gas phase. If indeed this cluster is stable in the gas phase, it would then be interesting to investigate the rotational freedom of the organic cation in the conformational space. An immediate follow-up question would intriguingly then be to elucidate the nature of the sole structural motif that is entirely responsible to provide incredible stability to the overall equilibrium geometry of the resulting binary cluster, as well as to infer what is responsible for the observed octahedral tilting discussed in the literature for the triiodide perovskite system in 3D.[30a),c)] Note that answers to these questions were partially yet indirectly provided by other authors in the past in studies that involved bulk calculation and others, but it is only just recently that a group of authors has pointed out that the exploration of the molecular characteristic properties of the small binary clusters as fundamental building blocks, such as CH$_3$NH$_3$PbI$_3$, CsPbI$_3$, and similar others, can be effective in understanding the fundamental cluster physics of materials of this kind.[21] Having knowledge of these, it would then be easier to explore more accurately the chemical physics associated with the molecular density of states, bandgaps, and optical energy gaps, of the extended perovskite systems in 2/3 dimensions. Specifically, giving examples, these authors[21] have shown that similar implementations were also carried out in the past for semiconductors such as Si, GaAs, PbS, ZnS, CdS, CdSe, and W$_2$O$_3$ to exploit their chemical physics. Such a thought-provoking implementation has greatly assisted in enlarging the scientific vision imaging big pictures from small clusters, the building blocks.

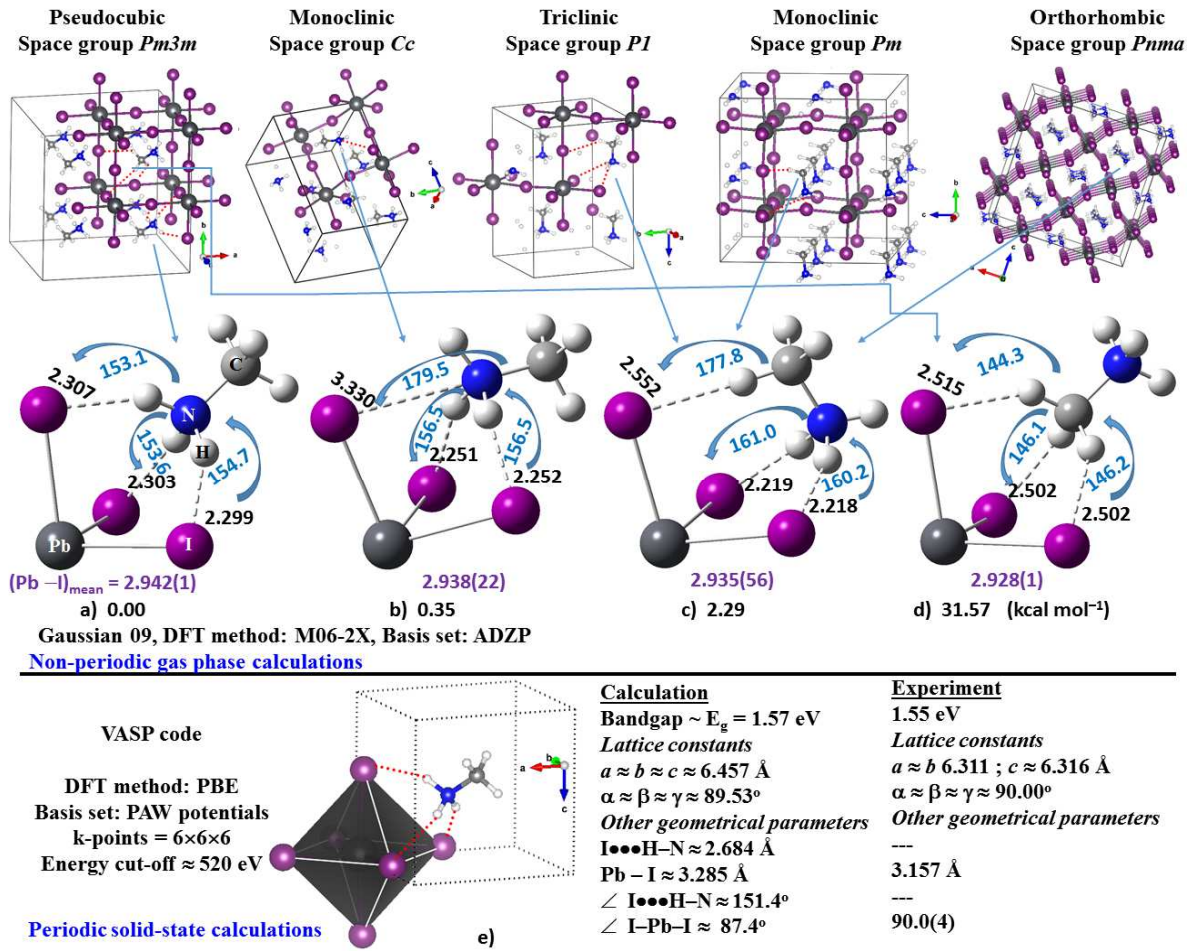

Fig. 1: The M06-2X/ADZP identified minimum-energy conformations of the $CH_3NH_3PbI_3$ nanocluster, a)–d). Illustrated are the relative energies, selected intermolecular bond distances (in Å) and bond angles (in ° respectively), and the mean Pb–I bond distances (in Å). e) The $CH_3NH_3PbI_3$ perovskite bulk, obtained using a spin-projected DFT calculation employing periodic boundary conditions. The solid lines between atoms in a)–e) represent covalent/coordinate bonds, while the dashed lines represent tentative intermolecular interactions (see text). Available experimental properties[16e] are also listed for comparison.

It was suggested that the physical/chemical stability of the $CH_3NH_3PbI_3$ perovskite stems from respective molecular motifs ($PbI_3^-$ and $CH_3NH_3^+$),[21] while others claim the role of the molecular cation is just only to dance.[60] This may mean it is hard to decide what is then the role of the organic cation? Note that the isolated ions, $PbI_3^-$ and $CH_3NH_3^+$, forming the $CH_3NH_3PbI_3$ system do not individually have any real physical existence during the absence of the other. In this respect, the properties of the CH$_3$NH$_3$PbI$_3$ perovskite solar cell emanates are collective. Otherwise, if these two isolated subunits individually can produce all sorts of functional optoelectronic properties, why then does one essentially invoke the organic part (CH$_3$NH$_3^+$) to couple it with the inorganic part (PbI$_3^-$) to demonstrate the chemical/physical stabilization of the entire CH$_3$NH$_3$PbI$_3$ supermolecular framework? The suggestion 'the organic cation only dance' is provided due to the assumed fact that it is perhaps difficult to accurately probe individual contributions of the two monomeric fragements when they are in the complex state, in which case, the molecular cation suppoosedly experiences dynamical motion (due to the effect of temperature) inside the space provided by the pseudocibic inorganic cage. No matter what happens, as already briefly outlined in the previous sections, we have delineated below in systematic detail using DFT calculations the results of the various geometrical, electronic, topological, and molecular orbital based properties of the CH$_3$NH$_3$PbI$_3$ molecular nanoclusters, which might greatly assist in the fundamental chemical and physical understanding of the 2D/3D frameworks of the CH$_3$NH$_3$PbI$_3$ perovskite solar cell.

Fig. 1 illustrates the four possible conformations of the CH$_3$NH$_3$PbI$_3$ building block, uncovered here for the first time with DFT-M06-2X/ADZP. These are all geometrically stable. They were resulted optimizing several different initial geometries of the lead triiodide perovskite block. All these (trial) initial geometries were prepared manually by rotating the CH$_3$NH$_3^+$ organic cation around the outer surface of the triangular I$_3^-$ face of the PbI$_3^-$ anion core, similarly as observed in the solid state geometries of the CH$_3$NH$_3$PbI$_3$ system experimentally. The triangular I$_3^-$ face is an octant of the corner-sharing PbI$_6^{4-}$ octahedron in the CH$_3$NH$_3$PbI$_3$ film in higher dimension.

The relative energy difference between the most and least stable conformers of the CH$_3$NH$_3$PbI$_3$ is approximately 31.57 kcal mol$^{-1}$ (see Rel. Stab. Column in Table 1 for details). The former gometry is analogous with its corresponding bulk (see lower panel in Fig. 1), predicted using periodic DFT calculation. This result is consistent with Leguy et al.,[32] who have used their quasielastic neutron scattering and molecular dynamics similation data to show that the CH$_3$NH$_3^+$ dipolar ions reorientate between the faces, corners or edges of the pseudocubic lattice cages in the CH$_3$NH$_3$PbI$_3$ crystals with a room temperature residence time of ~14 ps, and where free rotation, π-flips and ionic diffusion are ruled out within a 1–200-ps time window.

The four conformers of the CH$_3$NH$_3$PbI$_3$ cluster are not only stable in the gas phase at 0 K, but also thermodynamically stable at 298.15 K. The former feature is evident of the binding energies

$ΔE$ that are predicted to be negative for all the four conformers, while the latter is reminiscent of the changes in the Gibb's free energies $ΔG^0$ that are also large and negative (exothermic) for these conformers; both summarized in Table 1.

Table 1: Selected physical properties of the $CH_3NH_3^+•••^-I_3Pb$ nanoclusters, obtained with M06-2X/ADZP. [a]

| Type | Rel. Stab. | $ΔE$ | $ΔE(zpe)$ | $ΔG^0$ | I•••H–N | I•••H–C | cis-∠(I-M-I) | trans∠(I...H-N) | trans∠(I...H-C) | $E_g$[c] |
|---|---|---|---|---|---|---|---|---|---|---|
| Fig. 1a) | 0.00 | -125.11 | -125.06 | -114.24 | 2.303(4) | --- | 90.9(0) | 153.8(8) | --- | 6.86 |
| Fig. 1b) | 0.35 | -124.76 | -124.52 | -113.52 | 2.251(0) | 3.330[b] | 90.2(1) | 156.5(0) | --- | 6.79 |
| Fig. 1c) | 2.29 | -122.81 | -122.21 | -110.93 | 2.218(1) | 2.552 | 92.3(9) | 160.6(6) | 177.8 | 6.50 |
| Fig. 1d) | 31.57 | -93.53 | -92.08 | -81.64 | --- | 2.506(7) | 94.0(1) | --- | 145.5 | 4.75 |

[a] The properties include the relative stability (Rel. Stab. / kcal mol$^{-1}$), the binding energy ($ΔE$/kcal mol$^{-1}$), the zero-point corrected binding energy ($ΔE(zpe)$/kcal mol$^{-1}$), the change in the Gibb's free energy ($ΔG^0$/kcal mol$^{-1}$), the bond distances (r/Å) and selected bond angles (∠/°), and the HOMO-LUMO fundamental gap ($E_g$/eV).
[b] This distance corresponds to the I•••N(–C) intermolecular distance (see text for discussion), with ∠(I...N-C) ≈ 179.5°.
[c] Experimentally reported bandgap varies between 1.50 – 1.69 eV. [50,59]

Table 1 shows the magnitudes of $ΔE$ are unusually large for all the four conformers of the $CH_3NH_3PbI_3$ system. These are ca. –125.11, –124.76, –122.81, and –93.53 for conformers a), b), c) and d), respectively. Similarly, $ΔG^0$ are ca. –114.24, –113.52, –110.93, and –81.64 kcal mol$^{-1}$ for the corresponding conformers, respectively. It is also evident from Table 1 that the zero-point vibration has very small effect on the magnitude of $ΔE$ for the most stable conformer, but this not so for the other three (see $ΔE(zpe)$ of Table 1 for details). The very large binding energy is not the consequence of the basis set superposition error. In fact, our current investigation reveals almost similar magnitudes for $ΔE$ with CCSD(T) level thory in conjunction with triple-ζ quality pseudopotential basis sets such as cc-pVTZ (this calculation is carried out with a different research goal, and is beyond the scope of this article).

Fig. 2 presents the relaxed potential energy surface (PES) of the $CH_3NH_3^+•••^-PbI_3$ nanocluster, obtained using five correlated levels of theory. The minimum on the PES curve is appearing nearly around a Pb•••N distance of 3.9 Å, and the overall geometry belongs to it is the $CH_3NH_3PbI_3$ cluster illustrated in Fig. 1a. The PES curve is nearly a single well and asymmetric. There are two rudimentary regions on the PES curve, marked as a′ and a″. One is appearing in the 4.5 – 4.8 Å region, while the other is appearing in the 5.5 – 5.8 Å region. These two regions do not represent to any real local minima. Appearance of these is persistent regardless of the computational method applied. These appear due to the abrupt changes in the nature of the potential energy curve caused by the movement of the $CH_3NH_3^+$ species. The cluster at these two rudimentary positions shapes in such a way that the overall geometric configurations are unstable, attaining different modes of hydrogen binding interactions. Under any circumstance, the nature of the PES curve found with given a correlated method is identical with that found with the other. This may mean the energies of the intermolecular hydrogen bonding interactions involved to form the $CH_3NH_3^+\bullet\bullet\bullet^-PbI_3$ nanoclusters are a little affected by the nature of the correlated functionals when examined with the same basis set. Whether or not this is case is currently the subject of another benchmark study, and we will report it elsewhere. [33]

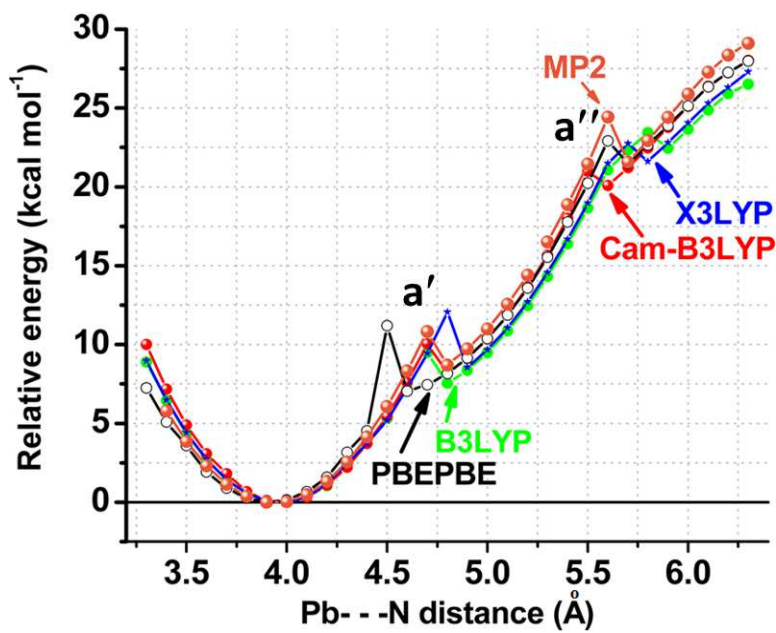

Fig. 2: Relaxed potential energy surface of the most stable $CH_3NH_3^+\bullet\bullet\bullet^-PbI_3$ nanocluster, obtained with five different correlated methods B3LYP, CAM-B3LYP, X3LYP, PBEPBE, and MP2(fc), where fc represents the frozen-core treatment and the DZP basis set. In this scan, the Pb•••N intermolecular distance was varied with step size of 0.1 Å for about 35 steps keeping all the other geometrical degrees of freedom to vary. The two rough regions marked as a′ and a″ on the PES curve correspond to abrupt changes in the overall geometry of the binary cluster that appear regardless of the nature of the correlated methods employed.

The unusually large $\Delta E$ calculated for the $CH_3NH_3PbI_3$ nanoclusters can be thought to arise largely from charge–charge Coulombic attraction because the two ions $PbI_3^-$ and $CH_3NH_3^+$ interacting with each other have opposite charges. In addition to this, charge–dipole and dipole–dipole electrostatic interactions are also expected to contribute prominently to $\Delta E$ of the clusters because $PbI_3^-$ and $CH_3NH_3^+$ possess dipole moments μ of 2.89 and 2.29 Debye, respectively. At the same time, polarization (or charge-transfer), and disperson are expected to contribute to $\Delta E$. This study does not undertake reporting a detailed investigation of separate component energies arising from the various interactions discussed above. Investigations to gain insight exactly into dissected energies responsible for explaining the intermolecular chemical-bonding interactions in the triiodide perovskite nanoclusters are currently underway, which will be reported elsewhere.[34]

As will be clearer later, our next question is: Do the I•••H short intermolecular distances illustrated in Fig. 1 for $CH_3NH_3PbI_3$ refer to hydrogen bonding[35] or halogen-bonding interactions?[36] In other words, do the short intermolecular contact distances in $CH_3NH_3PbI_3$ have any connection with any of the noncovalent interactions mentioned above, which are presumbly responsible for rendering the two subunits together in each of the four equilibrium configurations investigated? To answer a part of these questions, it is to be noted that the iodine atoms I involved in the formation of the I•••H intermolecular short contacts do not serve as electrophiles as they contain no regions of positive electrostatic potential on their outer surfaces in isolated $PbI_3^-$. The axial and equatorial outer surfaces of the iodine anion actually possess only negative electrostatic potentials, which are potentially involved in forming the intermolecular bonding interactions with positive sites of the organic species in one hand, and coordinate bonding interactions with the Pb divalent cation on the other, to form the $CH_3NH_3PbI_3$ cluster. Clearly, the presence of the I•••H motif in $CH_3NH_3PbI_3$ is not a consequence of halogen bonding,[36] as the formation of such an interaction may require the Pb-bound iodine to possess a region of positive potential on its outer electrostatic surface.

For characterizing whether the aforesaid interaction can be regarded as hydrogen bonding, one must concern with the question: what is a hydrogen bond? A hydrogen bond, symbolically represented with the motif A•••H–D, is simply a force of attraction between a positively charged covalently bound hydrogen and the negative site. In this motif the H atom is being shared between two electronegative species A and D (as in $H_2O$•••H–OH), which represent the hydrogen bond acceptor and donor species, respectively. Although a number of definitions (more than 50[37a]) exist in the vast literature starting from the first one provided by Huggins in 1921,[37a] it was only in 2009 that the IUPAC task group recommended a refined definition for it.[37b] It is accompanied by a list of six criteria, E1–E6, five characteristics, C1–C5, and nine footnotes, F1–F9; each may be evoked as a signature to identify the presence of hydrogen bonding interactions in molecules, in molecular clusters, and/or in solids. According to this latest definition, [37b] a typical hydrogen bond may be depicted as D–H•••A–Y, where the three dots denote the bond, and D–H represents the hydrogen bond donor. The acceptor may be an atom or an anion A, or a fragment or a molecule A–Y, where A is bonded covalently to Y. In specific cases, D and A can be the same with both D–H and A–H bonds being equal, with the acceptor as an electron-rich region such as, but not limited to, a lone pair in A or a π-bonded pair in A−Y. However, Desiraju, on the other hand, adviced that to define a hydrogen bond one must have a minimum of four atoms, D, H, A, and Y (or groups of atoms because a group of atoms can also constitute a valid acceptor fragment). The entire species D−H•••A−Y is properly considered as the hydrogen bond because each part (D−H, H•••A, and A−Y) affects the other parts and is likewise affected by them. It is misleading to think that only the H•••A part constitutes the hydrogen bond. To this end, let us consider the attractive interaction between X⁻ (X = I, Cl, Br) and H$_2$O, and trust on the the recommended guidelines of Desiraju as above.[35b] A question now aries, can we call the three dots in the X⁻•••H−O(H) motif as a hydrogen bond? The question is immediate because this motif does not include a minimum of four atoms to form the noncovalent interaction, nor one requires four atoms to define directionality in hydrogen bonds. In this context, a minimum of three aotms, such as X, H and O in the above motif, are more than sufficient to define a hydrogen bonding interaction. Even, one just needs two atoms (e.g. X and H in X⁻•••HOH) to identify the presence of a hydrogen bonding interaction between the neuclei of two bonded atomic basins.

We further comment that the above suggestion of Desiraju might be the case for systems with the H-bond donor and the H-bond acceptor containing both the fragments D−H and A−Y, but there are many occasions similarly as we showed above wherein one fails to fulfill the requirement Desiraju recommended for hydrogen bonding.[35b] Our oversimplified yet standalone adoption of the textbook level definition provided at the beginning of this paragraph is adequate to identify any strong hydrogen bonding interactions in chemical systems; again, this is regardless of whether the interaction is intermolecular or intramolecular.

Thus, following the oversimplified definition given above, it might be evident from Fig. 1 that the CH$_3$NH$_3$PbI$_3$ conformer a) conceives three I•••H−N hydrogen bonding interactions, while that of b) two I•••H−N, c) two I•••H−N and one I•••H−C, and d) three I•••H−C hydrogen bonding interactions, respectively. These interactions satisfy the *less than the sum of the van der Waals radii distance* criterion,[37b] in which there are significant mutual penetrations of the radii of both the I and H atoms upon formation of the I•••H hydrogenbonding interactions. The latter feature may indicate possibility of partial transfer of protons from the –NH$_3$ group (or from the –CH$_3$ group) in CH$_3$NH$_3$⁺ toward the inorganic cage nucleophiles upon the formation of perovskite films.

The nature of mutual penetration indicated above may be realized by looking at the mean I•••H(–C/N) intermolecular distances developed between the inorganic ($PbI_3^-$) and the organic ($CH_3NH_3^+$) fragments in the $CH_3NH_3PbI_3$ clusters (see Fig. 1 and Table 1 for values). For instance, the calculated mean I•••H(–C/N) distances are ca. 2.303, 2.251, 2.218, and 2.506 Å for the four conformers a), b), c), and d), respectively. Clearly, these each is less than the sum of the van der Waals radii, 3.24 Å of the H and I atoms (2.04 and 1.20 Å[38]), satisfying IUPAC recommended geometric criterion for identifying intermolecular hydrogen bonding interaction in chemical compounds.[37b)] The specific feature is not only true for each of the four gas phase clusters identified, but also evident in the geometry of the $CH_3NH_3PbI_3$ perovskite bulk (Fig. 1e)) (values 2.684 Å vs. 3.24 Å). The marked difference of the I•••H distances between the gas and bulk geometries suggests that such intermolecular hydrogen bonding interactions in $CH_3NH_3PbI_3$ may be stronger in the gas phase than in the solid state.[30a)–c)]

Are the hydrogen bonding interactions in $CH_3NH_3PbI_3$ directional? Perhaps yes. Three atoms (but not four) are generally used to define directionality in noncovalent interactions. In fact, an A•••H–D noncovalent interaction is said to be directional when the angle formed by the three atoms A, H and D (∠A•••H–D) involved directly in the formation of the intermolecular interaction lies in the range 160–180°.[35–37] In addition, there are discussions in the literature that suggest intermolecular interactions with the aforesaid angle in the 140–160° range can be regarded as a consequence of directional bonding. In the $CH_3NH_3PbI_3$ clusters, the ∠I•••H−N or the ∠I•••H−C angles formed by the three atoms I, H, and N (or I, H, and C) defining the intermolecular hydrogen bonding interaction can be used to define directionality. The mean values of ∠I•••H−N are ca. 153.8 ± 0.8°, 156.5 ± 0.0°, and 160.6 ± 0.6° for conformers a), b), and c), respectively, where ± refers to the standard deviations. Similarly, the mean values of the ∠I•••H−C are ca. 177.8 ± 0.0° and 145.5 ± 1.1° for conformers c) and d), respectively. From the QTAIM and NBO's second-order perturbation analysis presented below, we have identified another intermolecular interaction between an iodine atom in $PbI_3^-$ and the N atom of the –$NH_3$ group of $CH_3NH_3^+$ facing it in conformer b), that is, I•••N. This interaction is weak and directional. This is so because the intermolecular distance and the ∠I•••N−C associated with this interaction are ca. 3.330 Å and 179.5 ± 0.0°, respectively. These results signify the directional nature of all kinds of hydrogen bonding interactions identified in this study. This result suggests that directionality is indeed an important feature that might be considered in designing novel functional materials. We are currently investigating in detail the underlying importance of the hydrogen bonding interaction and directionality in the design principles of the 1D, 2D and 3D halide based perovskite systems. These supramolecular systems are the reinforced analogues of the presently studied nanoclusters depicted in Fig. 1, the physical chemistry of which will be reported elsewhere.

The massively large $\it{\Delta E}$ calculated for the four $CH_3NH_3PbI_3$ conformers are not comparable with strong-, medium- and weak-strength intermolecular interactions, as well as that with van der Waals.[35,23] For example, strong, medium, and weak inter/intramolecular interactions are proposed to have different ranges of values for the $\it{\Delta E}$.[23] Specifically, strong interactions comprise increasingly more covalent component than ionic with $\Delta E \geq -40$ kcal mol$^{-1}$, medium-strength interactions are largely electrostatic with partial covalency with $\Delta E \geq -15$ kcal mol$^{-1}$, weak-strength interactions are not just dispersive but include increasingly more electrostatic components with $\Delta E \geq -0.25$ kcal mol$^{-1}$, and those below the latter range are predominatly dispersive, which are mainly van der Waals. Obviously, the intermolecular hydrogen bonding interactions observed for the triiodide perovskite clusters are special, which are a kind of glue-like sticky adhesives that are essentially required for the rational design of novel soft nanomaterials. We classify these special intermolecular hydrogen bonding interactions as *ultra-strong* type, rather than calling them just as strongly/very strongly bound hydrogen bonding interactions. The classification is not very unsurprising because most of the characteristic properties proposed to date to identify and characterize any of the four types of noncovalent interactions described above do not appropriately fit with the properties of the intermolecular hydrogen bonding interactions uncovered for the conformers of the triiodide perovskite system.[23,35–37]

In a recent RSC article on chemical communication, Lee et al. discussed the results of their first-principles calculations on the orthorhombic phase of the $CH_3NH_3PbI_3$ solar cell system (see Scheme 1a) for graphical details).[30a)] According to them, relatively little progress has been made in the understanding of the role of hydrogen bonding in trihalide perovskites. So, they suggested their calculations have unraveled the presence of strong hydrogen bonding interactions in the $CH_3NH_3PbI_3$ crystal they exploited. Interestingly, though, they did not transmit any information on the quantitative strength of such interactions present in that system. Eventually, the authors concluded that these interactions not only influence the structure and dynamics of the $CH_3NH_3^+$ cation, but also reveal their liability for tilting of the PbI$_6^{4-}$ octahedra in the crystal. Because the N•••I distances were much shorter than those of C•••I and also the I•••H–C angle of approach was smaller than that of the I•••H–N angle, the authors supposed the hydrogen bond interactions mainly originate from H atoms on nitrogen. This is a statement which in our view is perversely counterintuitive since this is not the way how one may be tempted to define/characterize/identify a hydrogen bond in chemical systems (*vide supra*). T*he H atoms of CH$_3$NH$_3^+$ on carbon* are not involved in hydrogen bonding with the I atoms of the PbI$_6^{4-}$ octahedra is incorrect. Nor, such weak intereactions have any implication for lifting the stability of the geometry of the CH$_3$NH$_3$PbI$_3$ perovskite solar cell in either phases of the temperature is also incorrect. In addition, the authors of the same study have ignored all sorts of other intermolecular interactions, and have assumed that hydrogen bonding is important only when I•••H < 3 Å. (Why did the authors make such an assumption is unclear at this moment!) Based on this criterion Lee et al.[30a] could recognize only three I•••H–N hydrogen bonding interactions between the –NH$_3$ fragment of the CH$_3$NH$_3^+$ species and the Pb–I cubic cage interior that were proficient for the pronounced octahedral tilting observed in the orthorhombic geometry of the CH$_3$NH$_3$PbI$_3$ solar cell, although such a tilting in the low-temperature geometries of the system arises due to the presence of both interaction types, I•••H–N and I•••H–C.

Note that there are many scientific articles documented in the wide literature with demonstrations of weakly bound interactions > 3 Å (though very weakly!), which were masterful for the engineering designs of supramolecular materials of different kinds.[36,39] One such example might be found from a recent study of Kawai et al.,[40] wherein the authors introduced how tactfully can weakly bound F•••F interactions prompt the emergence of the supramolecular architecture they discovered on the Ag(111) surface.

Is it possible through any standard theoretical procedure to estimate *ΔE* of the individual hydrogen bonded interaction emanating from each of the four conformers of the CH$_3$NH$_3$PbI$_3$ illustrated in Fig. 1? No, it is not possible to estimate this energy accurately with the current state-of-the-art in modeling intermolecular interactions, such as with the supramolecular procedure we adopted in this study. However, because of the symmetrical nature of the geometries a) and d) (both nearly belong to the *C$_3$* point group), and that there are three nearly equivalent hydrogen bonds in each of them, the energy of each hydrogen bonding interaction in a) and that in d) can be crudely estimated by dividing the net *ΔE* of each nanocluster by three. The strength of each of such interactions is ca. −41.70 and −31.18 kcal mol$^{-1}$ in conformers a) and d), respectively. Howver, there is facilitation of another theoretical method, although very time consuming, which is an energy-decomposition procedure called the interacting quantum atoms (IQA) model.[41] This model can be used to assess the binding strength of each of the hydrogen bonding interactions in each of the $CH_3NH_3^+\bullet\bullet\bullet^-I_3Pb$ nanoclusters scrutinized in this study. While this procedure was originally proposed by Blanco et al.,[41a] application of this to study the detailed dissected energies of the $CH_3NH_3^+\bullet\bullet\bullet^-I_3Pb$ nanoclusters can be the topic of a future study.

Each of the three intermolecular bonding interactions in either of the four conformers of the $CH_3NH_3^+\bullet\bullet\bullet^-I_3Pb$, for example as in Fig. 1a), may not be treated as a separate interaction. One may understand it this way because the three intermolecular interactions in each conformer are collective, and should be treated as a single interaction. This is suggestive of the normal mode vibrational analysis, in which the formation of three intermolecular hydrogen bonding interactions in the $CH_3NH_3^+\bullet\bullet\bullet^-I_3Pb$ nanocluster resulted in only five normal modes of vibration. In general, the formation of a single yet linear intermolecular hydrogen bonding interaction results in one intermolecular stretching vibration, a degenerate in-plane bending vibration, and a degenerate out-of-plane deformation vibration.[42] While vibrational spectral studies of lead triiodide perovskite crystals have already been reported by other authors,[61] the underlying physical chemistry of molecular vibrations of the $CH_3NH_3^+\bullet\bullet\bullet^-I_3Pb$ nanoclusters may still be an important theoretical concern of a future investigation.

### 3.2  QTAIM and RDG-NCI descriptions of chemical bonding

Bader's QTAIM approach replies on the zero-flux boundary condition ($\nabla\rho(r_s).\hat{n}(r_s) = 0$, for for every point $r_s$ on the surface S($r_s$), $\hat{n}(r_s)$ is the unit vector normal to the surface at $r_s$, and $\rho$ is the charge density) is robust to unravel chemical bonding topologies in molecular and solid state systems.[25] The theory associates a set of four possible critical points of rank three called (3, –3) nuclear attractor critical point (NACP), (3, –1) bond critical point (BCP), (3, +1) ring critical point (RCP), and (3, +3) cage critical point (CCP), which satisfy the Poincaré–Hopf relationship, $n - b + r - c = 1$.[25a] In this relation, $n$, $b$, $r$, and $c$ are the number of NACPs, BCPs, RCPs, and CCPs, respectively, which are generally dictated from the gradient vector field of a charged density distribution that determines the molecular graph. It may be kept in mind that it is not necessary for every chemical system to have all four critical point types. This is so as the appearance of the ring and CCPs depends, among other things, on the nature of the charge distribution, the number of atoms involved, and the aromatic, nonaromatic, cyclic, and noncyclic natures of the chemical system examined. Thus, a diatomic molecule can be exemplified with two (3, –3) NACPs and one (3, –1) BCP. Similarly, an aromatic benzene molecule is exemplified with 12 (3, –3) NACPs, 12 (3, –1) BCPs, and 1 (3, +1) RCP, whereas the cubane molecule has 16 NACPs, 20 BCPs, 6 RCPs, and 1 CCP (see ref 51).

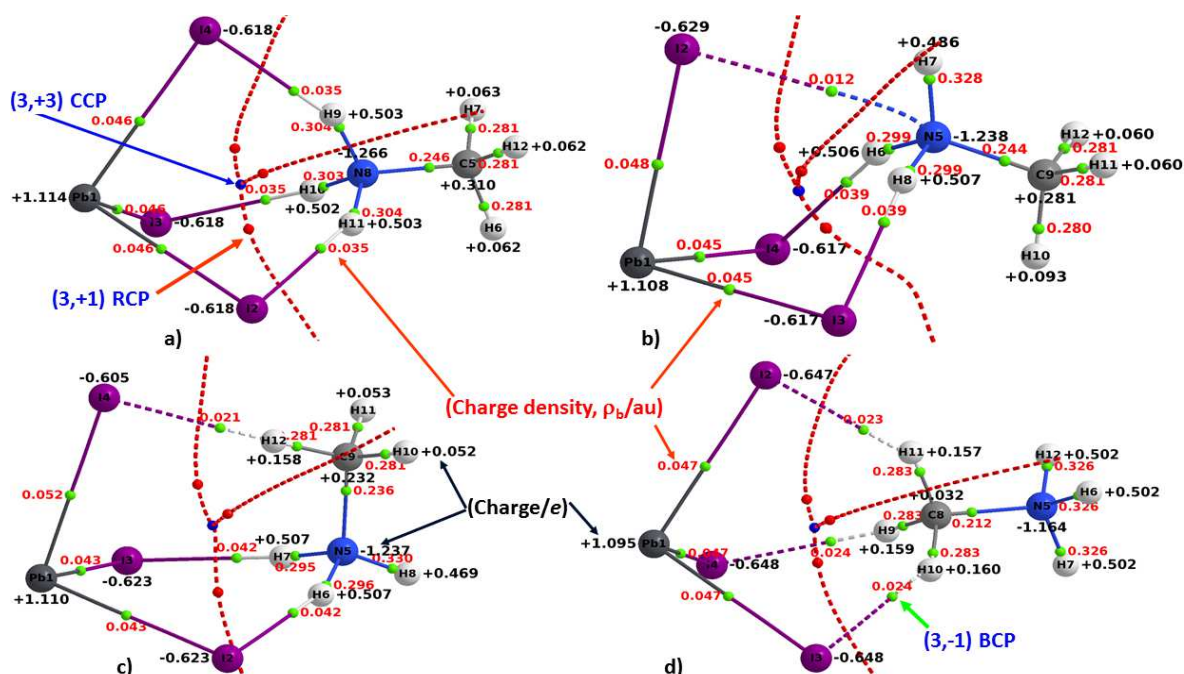

Fig. 3: M06-2X/ADZP level QTAIM molecular graphs for the $CH_3NH_3^+ \cdots {}^-PbI_3$ clusters. In a)–d), the iodine and lead atoms are displayed as large spheres in purple and dark gray, respectively, while the nitrogen and carbon atoms as medium spheres in blue and light gray, respectively, and the hydrogen atoms as tiny spheres in white-gray. The solid and dashed lines between atomic basins in atom color represent the bond paths, which each is a line of maximum charge density in accord with the connotation of Bader. The RCP attractor paths are illustrated by dashed lines in red. Tiny spheres in green, red, and blue represent the (3,–1) bond critical points (BCPs), (3,+1) ring critical points (RCPs), and (3,+3) cage critical points (CCPs), respectively. The charge density ($\rho_b$/au) at each bond critical point is shown in red, while the charges (in $e$) conferred on atoms are shown in black. Each molecular graph associates one CCP, three RCPs, and thirteen BCPs such that they all together satisfy the Poincaré–Hopf relationship, n + b + r + c = 1, where n, b, r, and c are the numbers of NACPs, BCPs, RCPs, and CCPs, respectively (see text).

Another descriptor of the theory is the bond path. It is the line of maximum charge density linking the nuclei of bonded atomic basins in chemical systems.[43a] Each bond path comprises a pair of trajectories originating at each (3, –1) critical point and terminates at the neighboring nuclear attractors. According to Bader,[43b] one may define a bond path operator as a Dirac observable, making the bond path the measurable expectation value of a quantum mechanical operator. One should not, therefore, treat a bond path as a chemical bond as it is only a valuable descriptor to represent the presence of any bonded interaction. While Pendás and collaborators introduced bond paths as exchange channels,[44a)] Tognetti and Joubert demonstrated that the concept of exchange channels introduced by Pendás et al. provides a means of rationalizing and predicting the presence of BCPs, enhancing the physical meaning of bond paths.[44b),c)]

Application of this approach to the conformations of the $CH_3NH_3PbI_3$ perovskite block has resulted in the molecular graphs illustrated in Fig. 3. These graphs are appropriate chemical representations of the perovskite block that can readily be inferred from chemical intuition. It is worth stressing that the $CH_3NH_3PbI_3$ cluster is an example of an ideal chemical system even from a topological perspective. This is because each of the four conformations of this cluster displays a topology, which comprises an identical number of critical points of the charge density. Interestingly, regardless of the nature of the orientation of the organic cation around the triangular facial part of the $PbI_3^-$, the resulting clustered geometry associates three ring critical points RCPs (each a characteristic of a ring surface) in a closely packed manner so as to produce a cage-like structure characterized by a cage crirical point CCP.

Table S1 summarizes numerical estimates of the various topological descriptors of the charge density for characterizing the chemical bonding interactions present in the $CH_3NH_3^+ \bullet\bullet\bullet^- PbI_3$ conformers. These include the charge density itself ($\rho_b$), the three eigenvalues of the Hessian of the charge density matrix ($\lambda_{i\,=\,1–3}$), the ellipticity ($\varepsilon_b = \lambda_2/\lambda_1 - 1$), the Laplacian of the charge density ($\nabla^2\rho_b = \lambda_1 + \lambda_2 + \lambda_3$, $\lambda_1 < \lambda_2 < \lambda_3$), the kinetic ($G_b$), potential ($V_b$), and total energy densities ($H_b$, $H_b = G_b + V_b$). From the data in the table, as well as from Fig. 3, it is obvious that $\rho_b$ lies in the 0.021–0.042 au range for the I•••H−N and I•••H−C hydrogen bonding interactions, with the former are stonger than the latter. For the I•••N−C, see b) of Fig. 3, the $\rho_b$ value is 0.012 au, showing that the former interactions are about two to three times stronger than this interaction. The $\rho_b$ for all these interactions are found to be smaller than the corresponding 0.043–0.052 au range calculated for the Pb–I coordinate bonds in $CH_3NH_3PbI_3$. The feature is not very surprising because the latter interactions are more covalent and ionic than the former interactions. The specific attribute can be grasped looking at the signs and magnitudes of both $\nabla^2\rho_b$ and $H_b$. For example, except for the I•••N−C, all other aforesaid interactions are characterized by $\nabla^2\rho_b > 0$ and $H_b > 0$. The former characteristic is indicative of the presence of significant involvement of ionic interactions, while the latter characteristic is indicative of the presence of covalency in all such interactions.[25a),43a)] Covalency both in hydrogen and coordinate-bonding interactions is generally expected when the potential energy density $V_b$ relative to the gradient kinetic energy density $G_b$ of the electrons dominates the total energy $H_b$ in those regions of space where the electronic charge is concentrated.[25a),43a)] This is possible especially when $G_b > 0$, $V_b < 0$, $|G_b| << |V_b|$, $H_b < 0$. The peculiarity is evident in the presently studied I•••H−N and I•••H−C hydrogen bonding interactions discussed above. While both the hydrogen and coordinate-bonding interactions in the CH$_3$NH$_3$PbI$_3$ clusters are a mixed type, our above claim that the latter interactions are more ionic and covalent than the former ones can be inferred from the signs and magnitudes of both $\nabla^2\rho_b$ and $H_b$ (for values see Table S1). Fig. 4 illustrates, as an example, both the contour plot and relief map of the Laplacian of the charge density, illustrating the regions of charge concentration and depletion in the most stable cluster of the CH$_3$NH$_3$PbI$_3$ perovskite block.

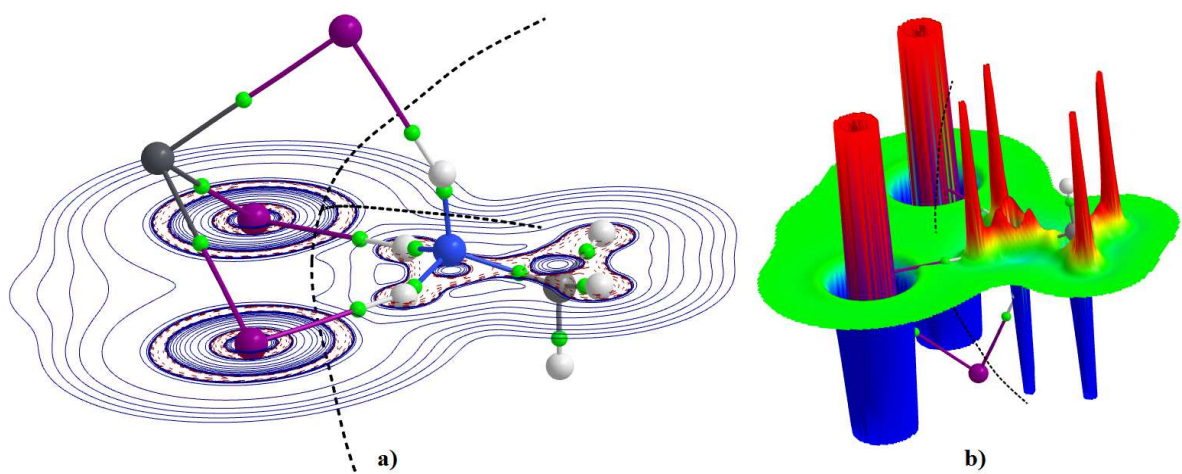

Fig. 4: a) 2D contour plot of the Laplacian of the charge density for the most stable CH$_3$NH$_3$$^+$•••$^-$PbI$_3$ nanocluster (see Fig. 1 for conformer type), obtained in the I–I–H plane. The dashed contour lines in red and solid lines in blue represent regions of charge concentration ($\nabla\rho_b < 0$) and depletion ($\nabla\rho_b > 0$), respectively. b) The relief map of the Laplacian of the charge density, truncated at a maximum height of 7.0 au.

For the I•••N−C interaction, see conformer b) of Fig. 3, $\rho_b \approx 0.0124$ au, $\nabla^2\rho_b \approx 0.0484$ au and $H_b \approx 0.0021$ au. Since $\nabla^2\rho_b$ and $H_b$ are both positive (i.e., $\nabla^2\rho_b > 0$ and $H_b > 0$) for this interaction, it is indifferent in characteristics from the hydrogen bonded interactions outlined above, and is predominantly electrostatic. The positive $\nabla^2\rho$ is mainly due to the eigenvalue $\lambda_3$ along the bond path direction, which is positive and substantial compared to those of the other two eigenvalues $\lambda_1$ and $\lambda_2$ in the other two perpendicular directions (see Table S1 for numerical details).

The charge density for the ring and cage structures characterized by the RCPs and CCPs are calculated to be very small. As expected, and for a given conformer, the order of ρ between the bond, ring and cage critical points is found to follow the conventional trend, that is, $\rho_b > \rho_r > \rho_c$, which is true irrespective of the nature of the conformations examined. A similar trend is found

for each of the properties $\nabla^2\rho$, G, V, and H, with the signs of the former two and latter one are all positive, and that for the remaining one is negative at RCPs and CCPs ($\nabla^2\rho > 0$ and H > 0), as expected. [25a),43a)]

As Jonson et al. demonstrated recently,[26a)] which is within the framework of the RDG-NCI theory, the sign of $\lambda_2$ determines the nature of an interaction and the value of $\rho$ determines the strength. This might not be very astonishing because at a nuclear position, $\rho$ is a maximum and the eigenvalues $\lambda_{i=1-3}$ are all negative, while at the cage or ring center $\rho$ is a minimum and the eigenvalues $\lambda_{i=1-3}$ are all positive. The remaining regions of space are characterized by a negative $\lambda_1$ ($\lambda_1 < 0$), a positive $\lambda_3$ ($\lambda_3 > 0$), and in this case, the sign of the second greatest eigenvalue $\lambda_2$ can either be positive ($\lambda_2 > 0$) or negative ($\lambda_2 < 0$). This is perhaps the reason the sign of $\lambda_2$ is a potential indicator to discriminate attractive interactions ($\lambda_2$ negative ($\lambda_2 < 0$)) from repulsive ones ($\lambda_2$ positive ($\lambda_2 > 0$)). Moreover, the RDG $s(r)$ exploits the chemical physics of NCI contained in Eqn 2, which describes the relationship between RDG $s(r)$, the charge density $\rho$, and its gradient $\nabla\rho$.[26]

$$RDG(r) = s(r) = \frac{1}{2(3\pi^2)^{\frac{1}{3}}} \frac{|\nabla\rho(r)|}{\rho(r)^{\frac{4}{3}}} \quad \ldots\ldots\ldots\ldots\ldots\ldots\ldots\ldots\ldots\ldots\ldots\ldots\ldots 2)$$

Fig. 5 illustrates the 2D plot of $s$ against $\rho \times$ sign ($\lambda_2$) that uncovers the repulsive and attractive (noncovalent) interactions the $CH_3NH_3^+\bullet\bullet\bullet^-I_3Pb$ chemical system comprises. With the help of the corresponding 3D isosurface plot that this approach emerges, the nature of these interactions can be vidualized. These are shown as the top and bottom graphics in each panel, respectively. The disc-like surfaces in deep-blue between the $Pb^{2+}$ and $I^-$ ions in a) indicate the charge density is largely accumulated in the bonding regions, while those between H atoms of the $-NH_3$ fragment in the organic cation and the $I^-$ ions are also accompanied by large concentration of charge density. As expected, the corresponding disc-like isosurfaces for the latter interactions are not markedly blue compared to those feasible for the coordinate interactions because of the appreciable difference in the charge density.

It is apparent from Fig. 5 that I•••H−C hydrogen bonding interactions in c) and d), as well as the I•••N−C noncovalent interaction in b), are all characterized by green-like isosufraces, with the latter one more faint. This result indicates that the charge density at the bonding regions of these interactions are significantly depleted compared to those found for the I•••H–N hydrogen bonding interactions viable in a), b) and c). All the features are also evident of their corresponding $s$ against $\rho \times$ sign ($\lambda_2$) 2D plots in the $\lambda_2 < 0$ regions, which appear as spikes. The locations of these spikes refer to the strengths of the intermolecular interactions revaled. These results clearly demonstrate that the chemical bonding topologies unraveled within the supermolecular framework of each conformer of the CH$_3$NH$_3$PbI$_3$ using QTAIM are in decent agreement with those originated from the RDG-NCI.

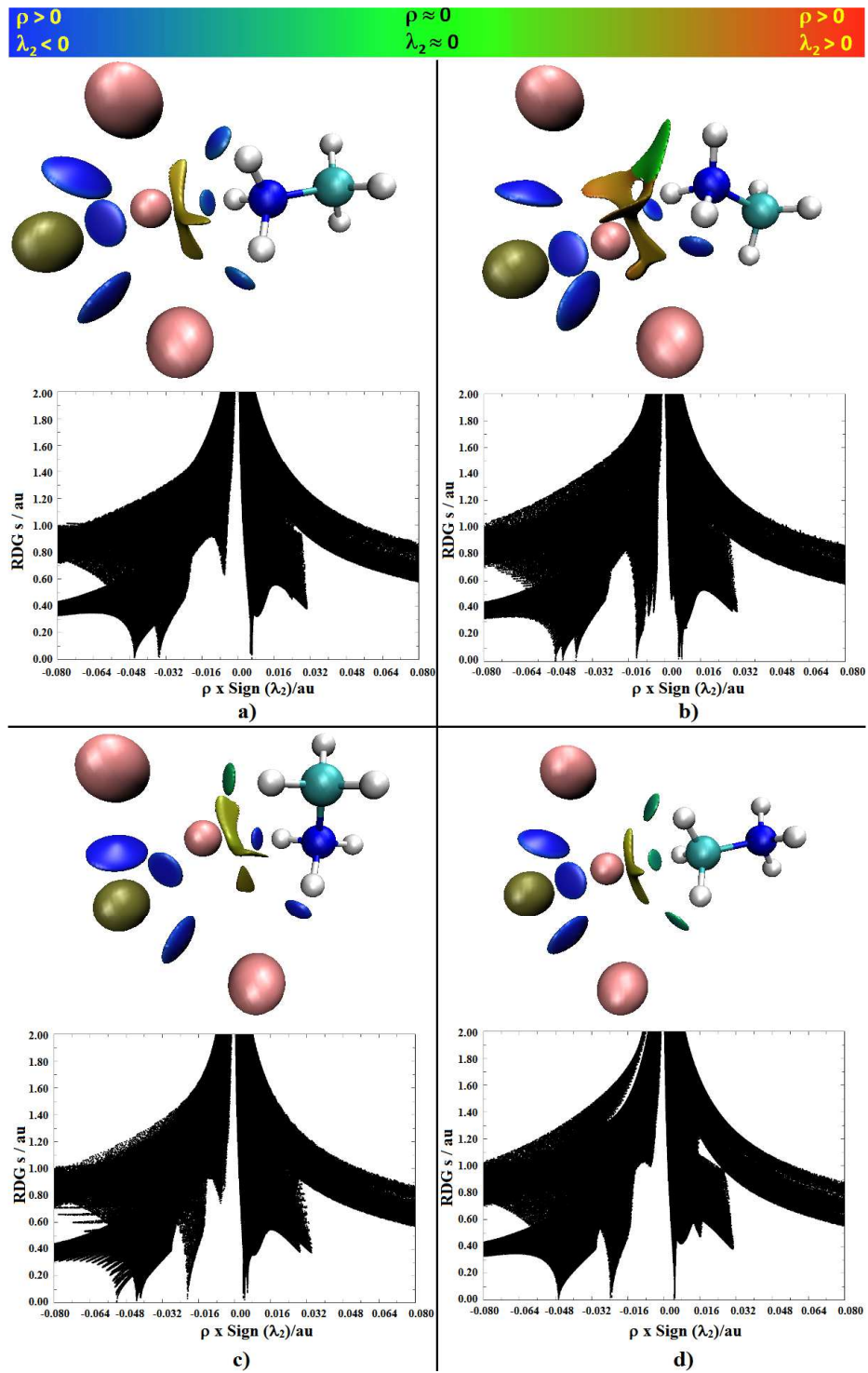

Fig. 5: The M06-2X/ADZP level RDG 0.3 au isosurface plots in 3D (top) and the s against $\rho \times$ sign ($\lambda_2$) plots 2D (bottom) for each conformer of the $CH_3NH_3^+$•••$^-I_3Pb$ nanocluster.

Table S1 also collects the the delocalization index ($\delta$), a measure of bond order,[52-53] for the most important atom-atom pairs. These pairs are already identified as intermolecular noncovalent interactions within the computational frameworks of QTAIM and RGD-NCI (see above). As can be seen from the data, Pb—Is′ are endowed with substantially large values of $\delta$, with values between 0.66 – 0.75. This gives further indication that the bonding interactions between the divalent metal ion and iodide anions do not involve any kind of $\pi$ character, as are purely singular bonds with significant mixing of covalent and ionic characters. On the other hand, $\delta$ for the I•••H(−N) interactions in a), b), and c) are ca. 0.15, 0.16, 0.18, respectively, showing their increasing order of their respective bonding stabilities. Similarly, $\delta$ for the I•••H(−C) interactions attainable in c) and d) are ca. 0.11 and 0.13, respectively. This also shows an increasing tendency for the strengthening of the intermolecular interaction passing from the former to the latter conformer. For the I•••N(−C) interaction realizable in b), $\delta$ is is found to be the smallest of 0.10. All these latter two types of intermolecular noncovalent interactions are weaker than the ones found in a), b), and c), as expected. For comparision, we cite Wang et al.,[54] who have previously reported comparable $\delta$ values of 0.192, 0.194, 0.196, 0.198, 0.199 for Li–H with HF, MP2, MP4, CISD, and QCISD, respectively. Notwithstanding the I•••N−C interaction uncovered in this study is a new type of noncovalent interaction, wherein two purely negative sites I and N interact constructively due to the influence of adjacent hydrogens (see further discussions given below). Studies on intermolecular interactions of this type may be discussed elsewhere.[36c-e),56]

### 3.3 Charge-rearrangement and –transfer

The utility of partial atomic charge to understand various chemical phenomena is very constructive, [36, 45a)-c)] even though some scientists question on its reliability.[45d),57] Nonetheless, partial atomic charge can be experimentally determined directly or indirectly, but also easily quantified theoretically using various population schemes, e.g., Mulliken, Hirshfield, QTAIM, NPA, CHELPG, and APT etc.[46]

The rearrangement of partial charges on atoms is a common feature that was seen in the past to assist in the fundamental understanding of the formation of supramolecular complex systems.[47] The same feature may be evident for the $CH_3NH_3^+\bullet\bullet\bullet^-I_3Pb$ clusters. To exploit the viability of this feature, it should be kept in mind that the fragment $PbI_3^-$ in its isolated form carries a net electronic charge of $-1.0$, and the $CH_3NH_3^+$ fragment a net electronic charge of $+1.0$. However, it is found that the formation of each of the four conformers of the $CH_3NH_3^+\bullet\bullet\bullet^-I_3Pb$ is accompanied with a significant rearrangement of partial atomic charges in the monomeric fragments. For example, as illustrated in Fig. 6 for the most stable conformer (true for the other three conformers), the $-1.0$ electronic charge on $PbI_3^-$ is increased to $-0.74$, and the $+1.0$ of electronic charge on $CH_3NH_3^+$ is decreased to $+0.74$, obtain using QTAIM. This means the net electronic charge on the $PbI_3^-$ fragment in $CH_3NH_3^+\bullet\bullet\bullet^-I_3Pb$ is progressively becoming more positive, while that on the $CH_3NH_3^+$ subunit is becoming less positive, both compared with the net charges conferred on each isolated counterpart. Similar results were also obtained for these conformers with the Hirshfield population scheme (values not given).[55]

Table 3: M06-2X/ADZP estimates of charge transfer ($\Delta Q/e$) between the fragments $PbI_3^-$ and $CH_3NH_3^+$ in the $CH_3NH_3^+\bullet\bullet\bullet^-I_3Pb$ nanoclusters.[a]

| Fig. 3 | $\Delta Q$(QTAIM) | $\Delta Q$(Hirshfield) |
|---|---|---|
| a) | 0.260 | 0.463 |
| b) | 0.246 | 0.429 |
| c) | 0.259 | 0.455 |
| d) | 0.151 | 0.270 |

[a] See Fig. 6 for the direction of charge transfer

Fig. 6: Schematic diagram for the transfer of charge $\Delta Q$ between $PbI_3^-$ and $CH_3NH_3^+$ upon the formation of the $CH_3NH_3^+\bullet\bullet\bullet^-I_3Pb$ cluster, with the black arrow indicates the direction of charge transfer. Shown are the QTAIM charges (in $e$) on atoms in isolated $PbI_3^-$ and $CH_3NH_3^+$. The formation of the $CH_3NH_3^+\bullet\bullet\bullet^-I_3Pb$ cluster leads the net $-e$ charge conferred on the $[PbI_3]^{-1}$ to increase (i.e., becoming more positive) and the net $+e$ charge $[CH_3NH_3]^{+1}$ to decrease (i.e., becoming less positive).

The rearrangement of atomic charges in molecular domains (when they are in the close proximity) is a consequence of electrostatic polarization. The extent of this polarization might be regarded as charge transfer ($\Delta Q$). Note that the concept of charge transfer is being increasingly recognized by scientists in the area of noncovalent interactions that include hydrogen bonding, halogen bonding, beryllium bonding, lithium bonding, chalcoegn bonding, σ$_{hole}$ interactions, and others. According to Clark and coworkers, the charge transfer theory is a mathematical attempt to represent a physical process, as it concerns with the mutual polarization of the interacting molecules, the so-called 'donor' and 'acceptor'. Thus it is redundant to cite both charge transfer and polarization as separate factors in noncovalent interactions.[57-58]

The ΔQ between PbI$_3^-$ and CH$_3$NH$_3^+$ is calculated be the largest for conformer a) compared to the three high-energy conformers (see Fig. 3 for conformer type and Table 3 for values). That is, ΔQ is the largest of 0.26 $e$ for the most stable conformer, and the least of 0.15 $e$ for the least stable conformer of the CH$_3$NH$_3^+$•••$^-$I$_3$Pb. In each case, it occurs in a direction from the PbI$_3^-$ fragment to the CH$_3$NH$_3^+$ fragment; this is analogously how the electronic charge is transferred from $^{11}$Na to $^{17}$Cl to form the Na$^+$Cl$^-$ salt. The trend in ΔQ by some means, though not very perfect, shows a close match with the $\Delta E$ for the CH$_3$NH$_3^+$•••$^-$I$_3$Pb clusters. The nature of charge transfer observed in this study is qualitatively is in good agreement with a very recent study of Madjet et al.[45c)]

### 3.4 The HOMO-LUMO fundamental gap

The difference between the energies of the HOMO (highest occupied molecular orbital) and the LUMO (lowest unoccupied molecular orbital) is called the HOMO-LUMO fundamental gap, denoted by the symbol $E_g$. Its value is tabulated in the last column of Table 1 for all the four conformers of the CH$_3$NH$_3^+$•••$^-$I$_3$Pb. The data show that $Eg$ is the largest of 6.86 eV for the most stable conformer a), in which case, the –NH$_3$ group in CH$_3$NH$_3^+$ is facing an octant in PbI$_6^{4-}$, and is the smallest of 4.75 eV for the least stable conformer, in which case, the –CH$_3$ group in CH$_3$NH$_3^+$ is facing an octant in PbI$_6^{4-}$. Evidently, $E_g$ for all the conformers are largely overestimated when they each is compared with the experimental $E_g$ range of 1.50 – 1.69 eV, [50] reported for the three phases of the CH$_3$NH$_3$PbI$_3$ perovskite solar cell (see Scheme 1). However, for the corresponding bulk displayed in Fig. 1e) an $E_g$ of ~1.57 eV is evaluated using periodic DFT calculation with PBE, reflecting an excellent agreement with experiment. While the gas phase calculation has predicted very large values for E$_g$, the HOMO-LUMO fundamental gap,[50,33] compared to experiment, it clarifies fact that the E$_g$ is not independent of the orientational disorder of the organic cation. This large deviation in the values of $E_g$ between the gas phase and experiment is not just due to the nature of the DFT functional employed, but also a solid state effect. Nonetheless, the large $E_g$ predicted with M06-2X/ADZP can be greatly reduced changing that DFT functional that carries 0% Hartree-Fock exchange. Also, the use of an appropriate basis set can reduce $E_g$ to a reasonable degree. These latter two conclusions are gleaned from a benchmark study that we are currently undertaking for the $CH_3NH_3^+ \bullet\bullet\bullet ^-I_3Pb$ nanoclusters, [33] which will report elsewhere.

### 3.5    NBO's second order perturbation theory analysis

According to second order perturbation theory analysis of Fock matrix in NBO basis,[24a] the formation of the $CH_3NH_3^+ \bullet\bullet\bullet ^-I_3Pb$ cluster may be understood as a result of hyperconjugation (charge transfer delocalization) between the $PbI_3^-$ and $CH_3NH_3^+$ species. As expected, this analysis dictates predominant charge transfer delocalization between the interacting monomers, see Table S2. For the most stable conformer a), it is occuring from the lone-pair (LP) electron donating orbitals of the three iodide atoms of the $PbI_3^-$ subunit (i.e., LP(3) I in $PbI_3^-$) to the σ* antibonding orbitals of the N–H bond(s) (i.e., σ*( N–H)) of the $CH_3NH_3^+$ subunit), i.e., LP(3) I → σ*(N–H), where (3) denotes to the third LP orbital. For conformation d), such predominant charge transfer interactions occurr between the LP of the electron donating orbitals of the three iodine atoms in $PbI_3^-$ and the σ* antibonding orbitals of the C–H bond(s) of the $CH_3NH_3^+$ subunit, i.e., LP(3) I → σ*(C–H). The strengths of the respective hyperconjugative interactions quantified with $E^{(2)}$ are ca. 40.0 and 15.0 kcal mol$^{-1}$ for conformations a) and d), respectively. Additional hyperconjugative interactions between various bonding and antibonding orbitals associated with the $PbI_3^-$ and $CH_3NH_3^+$ species are detailed in Table S2. These latter ones are having $E^{(2)}$ > 2.0 kcal mol$^{-1}$ in a) and ≈ 0.50 kcal mol$^{-1}$ in d), which include the [σ(Pb—I) → σ*(N–H)] and [σ(Pb—I) → σ*(C–H)] delocalizations, respectively, showing the involvement of the σ-type Pb—I coordination bonds in hyperconjugation. All these delocalizations are collectively liable for strengthening the stabilities of the $CH_3NH_3^+ \bullet\bullet\bullet ^-I_3Pb$ cluster.

For conformation b), two types of hypercojunctive interactions are evident. In it, one of them is LP(3) I → σ*(N–H) (for I•••H–N), which is relatively stronger than the weak one LP(3) I2 → σ*(1)N5—C9 (for I•••N–C), with $E^{(2)}$ for these 45.62 *vs.* 0.91 kcal mol$^{-1}$ (see Table S2). Interestingly, all these interactions were identified with QTAIM and RDG-NCI methods (*vide supra*). However, the former method predicts the latter interaction to be accompanied with a curved bond path (strained), and the latter method identified it through a green isosurface appeared between the two nuclei of bonded atomic basins. The attraction between the I and N atoms that leads to development of the I•••N−C is due to the joint influence of three adjacent hydrogen atoms in –NH$_3$ of CH$_3$NH$_3^+$ that show hyperconjgative interactions with the same iodide atom. Such hyperconjgations are described by LP (3) I2 → σ*(1) N5—H6, LP (3) I2 → σ*(1) N5—H7, and LP (3) I2 → σ*(1) N5—H8, with E$^{(2)}$ for these three lie between 0.8 and 1.1 kcal mol$^{-1}$.

For conformer c), the two types of hyperconjugative interactions are found, LP(3) I → σ*(N–H) and LP(3) I → σ*(C–H), which correspond to I•••H−N and I•••H−C, respectively, with the former being substantially weaker than the latter. This trend is evocative of E$^{(2)}$. For instance, E$^{(2)}$ amounts to 53.26 and 53.66 kcal mol$^{-1}$ for the two I•••H−N interactions, and that to 13.94 kcal mol$^{-1}$ for the I•••H−C (see Table S2 for details). In summary, it is to be concluded that the NBO results discussed above are in excellent agreement with those produced by QTAIM and RDG analyses, thereby recommending the fact that the employment of any of these three approaches can be advantageous in generating identical information about the topologies of chemical bonding interactions involved in these and analogous clusters of the trihalide perovskite.

Is the CH$_3$NH$_3^+$•••$^-$PbI$_3$ binary cluster a Mulliken inner complex? Perhaps yes. According to Mulliken and Person,[48] interactions based on the degree of charge transfer between the donor and acceptor molecules can be classified as Mulliken outer and inner complexes. While a typical hydrogen bond D•••H–A, predominantly of electrostatic origin, which is accompanied with little ΔQ, can be regarded as an outer complex, the inner complex is more strongly bound, and is generally identified in the form of an ion-pair, [D–H]$^-$•••A$^+$. According to Hill[19d] and Shaw et al.,[19c] the physical notion of an inner complex should not be taken as a requirement of a complete transfer of H$^+$, or X$^+$ (X = halogen derivative), but simply to signify significant charge transfer on complex formation. Typically, the ΔQ is less than 0.1e for weaker outer complexes, and those that are accompanied with large ΔQ can be regarded as Mulliken inner complexes. For example, thiirane•••ClF, NH$_4^+$•••$^-$Cl, and H$_3$P•••ICl are examples of Mulliken inner complexes because they involve ΔQ as large as 0.279, –1.0, and 0.144e, respectively. However, the thiirane•••HCl, oxirane•••HCl, NH$_3$•••Cl, H$_3$N•••XY (X, Y = halogen derivative) hydrogen bonded complexes are good examples of Mulliken outer complexes. Considering these observations as background, as well as the interpretations provided by others for different systems,[49] it is obvious that the CH$_3$NH$_3^+$•••$^-$PbI$_3$ perovskite clusters can be regarded as inner-type Mulliken complexes for several reasons. These include: i) the binding energies for the $CH_3NH_3^+ \bullet\bullet\bullet^- PbI_3$ clusters are unusually high, ii) they are complemented with very short $I \bullet\bullet\bullet H$ intermolecular distances of separation, and iii) they each is the consequence of a significant amount of transfer of charge between the interacting partners, $\Delta Q \gg 0.14\,e$, and iv) in agreement with iii), NBO's second-order perturbation analysis determined the very large charge transfer delocalizations ($E^{(2)}$ between 13.95 and 53.66 kcal mol$^{-1}$) that bring the two fragments together in clustered configurations.

### 4. Conclusions

In this study, we have performed M06-2X level DFT calculations using an all-electron double-$\xi$ ADZP basis set for the first time to study the geometrical and energetic stabilities, and the electronic and orbital properties of the fundamental building block of the numerously studied $CH_3NH_3PbI_3$ perovskite solar cell system in the gas phase. The non-periodic computational results obtained are compared with those reported experimentally in the solid state, as well with those that are obtained from this calculation employing periodic DFT. Among several other things, the study has:

i) Identified for the first time the possibility of four conformers for the $CH_3NH_3PbI_3$ solar cell system, which are very much likely to appear inside the inorganic cage with respect to the change in the phase of system. In two of them (the most and the least stable ones) the dipolar $CH_3NH_3^+$ ion is found to orientate toward the center of the face; in one it orients toward the corner, and in the other it parallels with the entire triangular facial part of the inorganic $PbI_3^-$ core. This result by some means is in appealing agreement with the recent results of Leguy *et al.* published in a Nature Communication article. [32]

ii) Uncovered for the first time the four conformers of the $CH_3NH_3^+ \bullet\bullet\bullet^- PbI_3$ nanocluster to have unusually large binding energies, which are about twice-thrice larger in magnitude than the covalent limit ($\sim -40$ kcal mol$^{-1}$). Such large binding energies are realized to be due to the presence of substantial covalency in the intermolecular hydrogen bonding interactions in each of the $CH_3NH_3^+ \bullet\bullet\bullet^- PbI_3$ perovskite nanoclusters.

iii) Showed the $CH_3NH_3^+ \bullet\bullet\bullet^- PbI_3$ perovskite nanoclusters to accompany very small intermolecular distances of separation, all much less than the sum of the van der Waals radii of the bonded atomic basins. Based on this attribute, together with ii), we classified the intermolecular hydrogen bonding interactions in $CH_3NH_3^+ \bullet\bullet\bullet^- PbI_3$ as *ultra-strong* type.

iv)    Discussed the occurrences of very large amount of charge transfers (> 0.14 $e$ and up to 0.26 $e$ with QTAIM) between the most important frontier occupied and unoccupied orbitals that have strongly participated in bringing the monomeric fragments $CH_3NH_3^+$ and $PbI_3^-$ together in complex configurations, a magnitude that is beyond the typical value < 0.1 $e$ observed for numerous Mulliken outer-type intermolecular hydrogen bonding interactions.

v)    Presented the fact that the role of the I•••C–H intermolecular hydrogen bonding interactions cannot/should not just be ignored/underestimated at any occasion, as was suggested previously. These interactions, simultaneously with the strongly bound I•••N–H hydrogen bonds, are found to be collectively responsible for the determinations of the stabilities and conformational/morphological freedoms of the $CH_3NH_3PbI_3$ solar cell in the solid state.

vi)    Revealed the fact that depending on the orientation of the organic cation around the facial part of the inorganic lead triiodide anionic species, the formation of attractive interactions between negative sites, such as the one discovered in this study, I•••N–C, is not very unlikely. Attractive interactions of this kind (i.e., between the negative sites) are already discussed by two of us previously. [36]

vii)    Utilized QTAIM and predicted the unified charge density topologies of chemical bonding interactions in the $CH_3NH_3^+$•••$^-PbI_3$ nanoclusters. In particular, each conformer is found to associate with similar number of BCPs, RCPs, and CCPs, and cage-like geometrical architectures. The bond path and BCP topologies that have faithfully uncovered the entire geometries of the $CH_3NH_3^+$•••$^-PbI_3$ nanoclusters are physically meaningful if visualized through chemical intuitions, as well as from the point of view of acid-base theory. QTAIM characterized the *ultra-strong* intermolecular hydrogen bonding interactions found in the $CH_3NH_3^+$•••$^-PbI_3$ nanoclusters as mixed type, meaning the intermolecular hydrogen bonding interactions are consequences of covalence and ionic interactions. The results of this theory are strongly supported by the 2D and 3D plots evolved from the RDG-NCI.

viii)    Employed NBO's second order perturbation analysis tool to insight into the various predominant channels for charge transfer delocalization, accompanying the formations of the $CH_3NH_3^+$•••$^-PbI_3$ nanoclusters. The results predicted showed that the $CH_3NH_3^+$•••$^-PbI_3$ nanoclusters examined are the consequences of the very large degrees of charge transfer delocalizations between various electron denoting and accepting bonding and anti-bonding orbitals, explained and quantified by hyperconjugative energies $E^{(2)}$.

Finally, based on the specialized characteristics presented above, especially ii), iii), iv), vii) and viii), the conformers of the $CH_3NH_3^+$•••$^-PbI_3$ perovskite solar cell block system are regarded as Mulliken inner complexes, a unique feature that probably has adequate consequence to design advanced functional nanomaterials and optoelectronic devices. This view may be in agreement with Hill *et al.*, and others,[19c)-d)] who have asserted that interactions of Mulliken inner type have desirable implications for fields such as crystal engineering and nanomaterials, where highly directional, strong interactions can be exploited for fine-control of the overall structure and properties. The results discussed here in light of the ultrastrong intermolecular hydrogen bonding interactions will greatly benefit researchers who are working on the future development of such highly valued functional materials for device based applications.

**Acknowledgement** All authors greatefylly acknowledge Institute of Molecular Science for the supercomputing computational facilities provided.

Organic-Inorganic Hybrid CH$_3$NH$_3$PbI$_3$ Perovskite Solar Cell Nanoclusters: Revealing Ultra-Strong Hydrogen Bonding and Mulliken Inner Complexes and Their Implication in Materials Design

Arpita Varadwaj,[*,a,b] Pradeep R. Varadwaj,[a,b] Koichi Yamashita[a,b]

[a]Department of Chemical System Engineering, School of Engineering, The University of Tokyo 7-3-1, Hongo, Bunkyo-ku, Japan 113-8656
[b]CREST-JST, 7 Gobancho, Chiyoda-ku, Tokyo, Japan 102-0076

# Supplimetary Information

[*]Corresponding Author's E-mail Addresses: pradeep@t.okayama-u.ac.jp (PRV); varadwaj.arpita@gmail.com (AV); yamasita@chemsys.t.u-tokyo.ac.jp (KY)

Table S1: Topological charge density properties of the $CH_3NH_3PbI_3$ nanoclusters, obtained with M06-2X/ADZP.[a,b]

| | Bond Type | $\rho_b$ | $\lambda_1$ | $\lambda_2$ | $\lambda_3$ | $\nabla^2\rho$ | $\varepsilon_b$ | $V_b$ | $G_b$ | $H_b$ | $\delta$ |
|---|---|---|---|---|---|---|---|---|---|---|---|
| a) | Pb1—I2 | 0.0460 | -0.0330 | -0.0329 | 0.1423 | 0.0764 | 0.01 | -0.0381 | 0.0286 | -0.0095 | 0.67 |
| | Pb1—I3 | 0.0459 | -0.0329 | -0.0327 | 0.1419 | 0.0763 | 0.01 | -0.0379 | 0.0285 | -0.0094 | 0.66 |
| | Pb1—I4 | 0.0460 | -0.0330 | -0.0328 | 0.1422 | 0.0764 | 0.01 | -0.0380 | 0.0286 | -0.0095 | 0.67 |
| | I2•••H11 | 0.0349 | -0.0382 | -0.0382 | 0.1351 | 0.0588 | 0.00 | -0.0234 | 0.0191 | -0.0044 | 0.15 |
| | I3•••H10 | 0.0354 | -0.0390 | -0.0389 | 0.1366 | 0.0587 | 0.00 | -0.0238 | 0.0193 | -0.0046 | 0.15 |
| | I4•••H9 | 0.0352 | -0.0386 | -0.0386 | 0.1360 | 0.0588 | 0.00 | -0.0237 | 0.0192 | -0.0045 | 0.15 |
| b) | Pb1—I2 | 0.0484 | -0.0349 | -0.0347 | 0.1461 | 0.0764 | 0.01 | -0.0408 | 0.0299 | -0.0108 | 0.70 |
| | Pb1—I3 | 0.0451 | -0.0324 | -0.0322 | 0.1410 | 0.0764 | 0.01 | -0.0372 | 0.0281 | -0.0091 | 0.65 |
| | Pb1—I4 | 0.0451 | -0.0324 | -0.0322 | 0.1410 | 0.0764 | 0.01 | -0.0372 | 0.0281 | -0.0091 | 0.65 |
| | I2•••N5 | 0.0124 | -0.0074 | -0.0031 | 0.0589 | 0.0484 | 1.39 | -0.0078 | 0.0100 | 0.0021 | 0.10 |
| | I3•••H8 | 0.0392 | -0.0448 | -0.0441 | 0.1478 | 0.0588 | 0.02 | -0.0272 | 0.0210 | -0.0062 | 0.16 |
| | I4•••H6 | 0.0392 | -0.0449 | -0.0442 | 0.1479 | 0.0588 | 0.02 | -0.0272 | 0.0210 | -0.0063 | 0.16 |
| c) | Pb1—I2 | 0.0435 | -0.0311 | -0.0308 | 0.1383 | 0.0764 | 0.01 | -0.0355 | 0.0273 | -0.0082 | 0.63 |
| | Pb1—I3 | 0.0434 | -0.0311 | -0.0307 | 0.1381 | 0.0763 | 0.01 | -0.0354 | 0.0272 | -0.0082 | 0.63 |
| | Pb1—I4 | 0.0522 | -0.0379 | -0.0375 | 0.1544 | 0.0790 | 0.01 | -0.0455 | 0.0327 | -0.0129 | 0.75 |
| | I2•••H6 | 0.0417 | -0.0487 | -0.0482 | 0.1548 | 0.0579 | 0.01 | -0.0293 | 0.0219 | -0.0074 | 0.18 |
| | I3•••H7 | 0.0418 | -0.0488 | -0.0483 | 0.1549 | 0.0578 | 0.01 | -0.0293 | 0.0219 | -0.0074 | 0.18 |
| | I4•••H12 | 0.0209 | -0.0200 | -0.0198 | 0.0898 | 0.0500 | 0.01 | -0.0129 | 0.0127 | -0.0002 | 0.11 |
| d) | Pb1—I2 | 0.0468 | -0.0336 | -0.0332 | 0.1445 | 0.0776 | 0.01 | -0.0392 | 0.0293 | -0.0099 | 0.68 |
| | Pb1—I3 | 0.0467 | -0.0335 | -0.0332 | 0.1443 | 0.0776 | 0.01 | -0.0391 | 0.0292 | -0.0099 | 0.68 |
| | Pb1—I4 | 0.0467 | -0.0336 | -0.0332 | 0.1444 | 0.0776 | 0.01 | -0.0392 | 0.0293 | -0.0099 | 0.68 |
| | I2•••H11 | 0.0232 | -0.0220 | -0.0213 | 0.0998 | 0.0565 | 0.03 | -0.0152 | 0.0147 | -0.0005 | 0.13 |
| | I3•••H10 | 0.0237 | -0.0226 | -0.0220 | 0.1016 | 0.0569 | 0.03 | -0.0156 | 0.0149 | -0.0007 | 0.13 |
| | I4•••H9 | 0.0237 | -0.0226 | -0.0220 | 0.1015 | 0.0569 | 0.03 | -0.0156 | 0.0149 | -0.0007 | 0.13 |

[a] See Fig. 1 and Table S2 for conformational and atomic labeling details.
[b] Properties include the charge density ($\rho_b$/au), the three principal eigenvalues of the charge density matrix ($\lambda_{i=1-3}$/au), the Laplacian of the charge density ($\nabla^2\rho$/au), the ellipticity ($\varepsilon_b$), the potential energy density ($V_b$/au), the kinetic energy density ($G_b$/au), total energy density ($H_b = V_b + G_b$/au), and the delocalization index ($\delta$). $\varepsilon_b$ and $\delta$ are unitless.

Table S2: Second order hyperconjugation interaction energies E$^{(2)}$ associated with various important electron delocalizations between the donor NBOs (i) and acceptor NBOs (j) for the CH$_3$NH$_3^+$•••$^-$PbI$_3$ clusters. [a]

| Donor NBO (i) | | Acceptor NBO (j) | E$^{(2)}$/ kcal mol$^{-1}$ |
|---|---|---|---|
| **Conformer a)** | | | |
| LP (3) I2 | → | σ*(1) N8—H11 | 38.57 |
| LP (3) I3 | → | σ*(1) N8—H10 | 39.95 |
| LP (3) I4 | → | σ*(1) N8—H9 | 39.28 |
| BD (1)Pb1—I2 | → | σ*(1) N8—H11 | 2.14 |
| BD (1)Pb1—I3 | → | σ*(1) N8—H10 | 2.43 |
| BD (1)Pb1—I4 | → | σ*(1) N8—H9 | 2.25 |
| **Conformer b)** | | | |
| LP (3) I2 | → | σ*(1) N5—H6 | 0.80 |
| LP (3) I2 | → | σ*(1) N5—H7 | 1.08 |
| LP (3) I2 | → | σ*(1) N5—H8 | 0.80 |
| LP (3) I2 | → | σ*(1) N5—C9 | 0.91 |
| LP (3) I3 | → | σ*(1) N5—H8 | 45.62 |
| LP (3) I4 | → | σ*(1) N5—H6 | 45.72 |
| LP (1) I3 | → | σ*(1) N5—H8 | 4.84 |
| LP (1) I4 | → | σ*(1) N5—H6 | 4.85 |
| BD(1)Pb1—I3 | → | σ*(1) N —H8 | 6.27 |
| BD(1)Pb1—I4 | → | σ*(1) N5—H6 | 6.29 |
| **Conformer c)** | | | |
| LP (3) I2 | → | σ*(1) N5—H6 | 53.26 |
| LP (1) I2 | → | σ*(1) N5—H6 | 5.02 |
| BD(1)Pb1—I2 | → | σ*(1) N5—H6 | 1.65 |
| LP (3)I3 | → | σ*(1) N5—H7 | 53.66 |
| LP (1) I3 | → | σ*(1) N5—H7 | 5.04 |
| BD(1)Pb1—I3 | → | σ*(1) N5—H7 | 1.67 |
| LP (3)I4 | → | σ*(1) C9—H12 | 13.94 |
| BD(1)Pb1—I4 | → | σ*(1) C9—H12 | 1.05 |
| **Conformer d)** | | | |
| LP (3) I2 | → | σ*(1) C8—H11 | 14.22 |
| BD (1)Pb1—I2 | → | σ*(1) C8—H11 | 0.49 |
| LP (3) I3 | → | σ*(1) C8—H10 | 15.16 |
| BD (1)Pb1—I3 | → | σ*(1) C8—H10 | 0.59 |
| LP (3)I4 | → | σ*(1) C8—H9 | 15.16 |
| BD (1)Pb1—I4 | → | σ*(1) C8—H9 | 0.60 |

[a] LP, BD, and σ*, represent the lone-pair, bonding, and anti-bonding natural orbitals involved, respectively.